\documentclass[11pt]{article}
\usepackage[T1]{fontenc}
\usepackage[utf8]{inputenc}
\usepackage[margin=1in]{geometry}
\usepackage[authoryear]{natbib}
\usepackage{graphicx}
\usepackage{amssymb,amsmath}
\usepackage{threeparttable}
\usepackage{placeins}
\usepackage{url}
\usepackage{hyperref}
\hypersetup{colorlinks=true,linkcolor=blue,citecolor=blue,urlcolor=blue}

\newcommand{\xb}[1]{\pmb{#1}}
\newcommand{\xbi}[1]{\pmb{\mathit{#1}}}
\newcommand{\xbu}[1]{\underline{\pmb{#1}}}
\newcommand{\xbiu}[1]{\underline{\pmb{\mathit{#1}}}}

\title{}
\author{}
\date{}

\begin{document}

\begin{center}
{\LARGE 	An AI-Ready Pipeline for Impedance-Resolved QCM Biosensor: Interpretable Line-Shape Features, Redundancy Control, and Robust Regression\par}
\vspace{1em}
Ceyhun K{\i}r{\i}ml{\i}$^{1,*}$, El\c{c}im Elg\"un$^{2}$, Selin Ya\u{g}mur Tu\u{g}ta\u{g}$^{1}$\par
\vspace{0.5em}
$^{1}$Ac{\i}badem Mehmet Ali Ayd{\i}nlar University, Department of Biomedical Engineering, Ata{\c{s}}ehir, Istanbul, T{\"u}rkiye\par
$^{2}$Ac{\i}badem Mehmet Ali Ayd{\i}nlar University, Department of Computer Engineering, Ata{\c{s}}ehir, Istanbul, T{\"u}rkiye\par
\vspace{0.25em}
$^{*}$Corresponding author\par
\end{center}

\begin{abstract}
Accurate inference from quartz crystal microbalance (QCM) measurements in liquids is often limited by reducing resonance behavior to two scalar endpoints (frequency and dissipation shifts, $\Delta f$ and $\Delta D$) or by relying on single-equation analytical models (Kanazawa-model). We propose an AI-ready, impedance-resolved workflow that preserves full resonance line-shape information and converts it into compact, physically interpretable features for supervised regression. A passive microfluidic mixer generates glycerol--water concentration gradients under a constant total flow rate (50~$\mu$L/min), while complex impedance spectra of a 10~MHz AT-cut quartz crystal are recorded in real time. Each sweep of nine spectra is parameterized via constrained Gaussian/Lorentzian models to yield 52 line-shape descriptors spanning extrema of $|Z|$, $X$, and $B$ and peaks of $R$, phase, and $G$. The pipeline integrates consensus outlier handling, redundancy-aware feature ranking (mRMR), and cross-validated regression across linear, kernel, and ensemble models. Compared with the classical Kanazawa baseline, the impedance line-shape approach reduces concentration-prediction error from 0.456 to 0.148~\%v/v RMSE (3.09$\times$ lower). The results demonstrate that impedance-resolved line-shape features provide a robust and interpretable basis for machine-learning-assisted QCM inference and illustrate a generalizable pattern for AI-enabled spectral sensing.
\end{abstract}


\section{Introduction}
\label{sec:introduction}

Quartz crystal microbalance (QCM) sensing offers real-time, label-free transduction and has become a workhorse for probing interfacial processes, adsorption, and viscoelastic films in liquids. In the rigid, thin-film limit, the resonance frequency shift $\Delta f$ follows Sauerbrey’s gravimetric relation \cite{Sauerbrey1959}. In liquid-phase operation and for soft or viscoelastic loads, the response reflects coupled inertial, viscous, and elastic contributions; consequently, interpretation relies on hydrodynamic and viscoelastic frameworks such as the Kanazawa--Gordon bulk-loading model \cite{KanazawaGordon19851,KanazawaGordon19852} and layered viscoelastic treatments (e.g., Voinova-type approaches) \cite{Voinova1999_PhysScr,Voinova2002_BiosensBioelectron}, with broader conceptual discussion by Johannsmann \cite{Johannsmann2006}. Modern practice and application breadth are summarized in recent QCM/QCM-D and acoustic-wave sensing reviews \cite{Easley2022_QCMD_Guide,TondaTuro2018_Frontiers_QCMD,Songkhla2021_Chemosensors_QCMBehavior,Johannsmann2021_Sensors_Review,Arnau2008_Sensors,Alassi2017_QCM_Interfacing,Huang2021_SAW_BiosensingReview,RochaGaso2009_SGAW_Review}.

Despite this strong physical foundation, many workflows still compress rich resonance information to a small set of scalar descriptors, most commonly $\Delta f$ and a dissipation- or bandwidth-related metric ($\Delta D$ or $\Delta\Gamma$). Such low-dimensional readouts are attractive for routine use but can be information-limiting for quantitative inference when multiple mechanisms (bulk viscosity/density changes, interfacial slip, viscoelastic relaxation, surface roughness, or mixed-mode loading) produce partially overlapping contributions \cite{Johannsmann2006,Arnau2008_Sensors,Alassi2017_QCM_Interfacing,Daikhin2002_AnalChem_AdmittanceRoughness}. Similar limitations arise when viscosity/density effects are inferred from a single closed-form relation (e.g., Kanazawa-type expressions), because the response depends on coupled fluid properties and can be perturbed by real-world experimental conditions; this motivates impedance-resolved, parameter-rich and regression-based inference approaches in complex liquids.
\cite{Liao2022_Sensors_BloodViscosity_QCM,Burda2022_Sensors_ImpedanceQCM,Voglhuber2019_QTF_ViscosityDensity}. This creates a clear opportunity for data-driven inference: QCM provides information-rich spectra, but standard $\Delta f/\Delta D$ readouts collapse them to low-dimensional signals, motivating ML pipelines that exploit the retained spectral structure while preserving physical interpretability.

A direct route to increased information utilization is impedance- (or admittance-) resolved analysis, where the full resonance line shape is acquired and parameterized rather than reduced to two metrics. Prior work has shown that full admittance/impedance spectra contain richer structure than a single frequency shift, and that multivariate analysis of spectral descriptors can separate different physical contributions in liquid-phase QCM measurements \cite{Chen1995_Analyst_PCA_Admittance,ButtryWard1992_ChemRev_EQCM,DeakinButtry1989_AnalChem_EQCMApplications,Huang2017_ResistanceAmplitude_QCM}. In particular, relating the complete admittance spectrum of a loaded resonator to material and interfacial properties has been shown to enable quantitative inference that is inaccessible to scalar reductions alone \cite{Gillissen2017_AnalChem_AdmittanceSpectrum}. In parallel, electrochemical impedance spectroscopy has seen growing adoption of machine-learning and multivariate approaches for automated and more robust inference in complex settings, spanning impedimetric biosensing and bioanalytical sensing as well as energy-storage prognostics under variable operating regimes \cite{Shimizu2023_TrAC_ML_ImpedimetricBiosensing,Randviir2022_AnalMethods_EIS_Bioanalytical,Doonyapisut2023_AIChEAI_EIS_DL,Jones2022_CellReportsPhys_BatteryEIS_ML,Zhang2025_Biosensors_EIS_PathogenReview}. These developments motivate transferring similar AI principles—robust preprocessing, redundancy control, and cross-validated model selection—to impedance-resolved QCM line-shape data.

In this work, we develop an impedance-resolved QCM inference framework that exploits full resonance line-shape information rather than reducing measurements to a two-scalar $\Delta f/\Delta D$ readout. The proposed pipeline is designed around the empirical characteristics of the acquired constant total flow dataset: descriptor redundancy is managed using mutual-information–based minimum redundancy maximum relevance (mRMR) ranking with top-$k$ ablations, while transient artifacts and occasional fit failures are mitigated through staged outlier screening and a soft-consensus ensemble filter. To capture nonlinear behavior without overfitting, we compare linear baselines and nonlinear regressors under a unified cross-validation protocol, and we retain interpretability by using physically grounded line-shape descriptors obtained from constrained Gaussian/Lorentzian parameterization rather than opaque latent embeddings. This framework builds on our group’s prior work on impedance-resolved QCM feature engineering and ML-centric analysis workflows \cite{Kirimli_2014,Kirimli2022_ML_Optimization,Kirimli2024_EDA,Kirimli_2025,Kirimli_2025_Conference}. We benchmark the approach against conventional $\Delta f/\Delta D$-style calibration and the Kanazawa baseline, demonstrating improved prediction performance and robustness across the studied concentration range. The remainder of the paper details the experimental/microfluidic setup and spectral parameterization, the feature processing and learning pipeline, and the comparative results and discussion.

\section{Materials and Methods}
\subsection{Fabrication of flow cell with passive mixer}
\subsubsection{CAD/CAM and machining}
The flow cell comprises three layers of 3-mm cast acrylic (PMMA). Geometry and toolpaths were prepared in Autodesk Fusion 360 and milled on an in-house built CNC. Mixer features were cut with a $0.5 mm$ flat end mill inside a 3D-printed water bath for chip removal and thermal control. Unless noted, the feed rate was $300 mm.min^{-1}$ and spindle speed $8000 rpm$. All ports, through-holes, and the QCM seat were machined before bonding.
\subsubsection{Passive mixer geometry}
The middle layer contains an array of circular baffles (pillars) to enhance transverse advection. Pillar diameter is $1.25 mm$; the gap between adjacent pillars is $\approx 0.53 mm$. The channel height is $0.50 mm$. These dimensions balance hydraulic resistance with mixing efficiency and respect the end-mill diameter.
\subsubsection{QCM sealing}
A 10 MHz plano-convex AT-cut QCM (vendor: openQCM) is sealed between two elastomer O-rings (major/minor radii $11.1/1.6 mm$) seated in concentric grooves. The crystal is clamped between the fused (top+middle) assembly and the bottom plate, with O-ring compression defining the wetted footprint and preventing bypass flow.
\subsubsection{Electrical interfacing}
A custom PCB carries two spring-loaded pogo pins that contact the QCM electrodes from below through clearance holes in the bottom plate and O-ring groove. The PCB mates to a Keysight E4990A via the 16047E fixture.
\subsubsection{Thermal bonding (heat fusing)}
The top and middle acrylic layers were permanently joined by controlled thermal bonding. To prevent warping under load, a snug MDF jig was CNC-cut to house the acrylic pair; the jig was clamped between flat aluminum plates, and binder clips provided uniform clamping pressure (layout provided in the Supplemental Information). The stack was heated in a box oven with a $2 ^{o}C.min^{-1}$ ramp to 115 °C, held 30 min, then cooled to room temperature under load, yielding a clear, void-free bond.
\subsubsection{Final assembly}
The bonded top–middle assembly, QCM, and bottom plate were stacked with the two O-rings and fastened using screws (Figure 1a). The PCB was then attached so the pogo pins engaged the electrodes. Figure 1a shows the assembled flow cell with the passive mixer; Figure 1b shows machined faces of the acrylic layers (top on the left, bottom on the right). The long pieces in Figure 1b correspond to the top and middle layers that were heat-fused. A step-by-step workflow, jig drawings, clamping pattern, and post-bond cleaning are provided in the Supplemental Information Section S.1.
\subsection{Flow setup}
Two inlet streams were driven by a dual-channel Elveflow OB1 MK3+ pressure controller equipped with inline flow sensors; each channel was regulated by proportional–integral (PI) feedback to enforce the programmed rates. Channel~1 (DI water) followed a sinusoidal profile between 30 and 50~$\mu$l/min with a 1~h period. Channel~2 (5\% v/v glycerol) was commanded with an equal-amplitude sinusoid phase-shifted by 180$^\circ$, varying between 0 and 20~$\mu$l/min. The phase offset maintained a constant total flow of 50~$\mu$l/min while modulating composition. Maintaining constant total flow is critical because QCM signals are sensitive to flow-induced hydrodynamic loading; fixing the flow rate reduces this source of variability. After merging, the baffle mixer produced a sinusoidal glycerol concentration at the sensor of 0--2\% (v/v). All solutions, the flow cell, and the flow sensors were connected using PTFE tubing (inner diameter 1.0~mm, outer diameter 1.5~mm) with polyether ether ketone (PEEK) fittings and adapters to ensure chemical compatibility and low adsorption. For stable operation of the three-port passive mixer (two inlets, one outlet), the OB1 supplied positive pressure at the inlets; when the PI loop required negative gauge pressure to meet a setpoint, an auxiliary vacuum line (via the OB1 vacuum port/external source) was applied to the inlets. The outlet remained at atmospheric pressure and was routed to a waste container, which helped prevent backflow and stabilized the junction while maintaining the programmed waveform. Flow-rate data sampled at 20~Hz were used to compute real-time concentrations and were interpolated to synchronize with the impedance-sweep timestamps. Figure1c shows the flow setup and components.

\begin{figure}
\centerline{\includegraphics[width=0.7\textwidth]{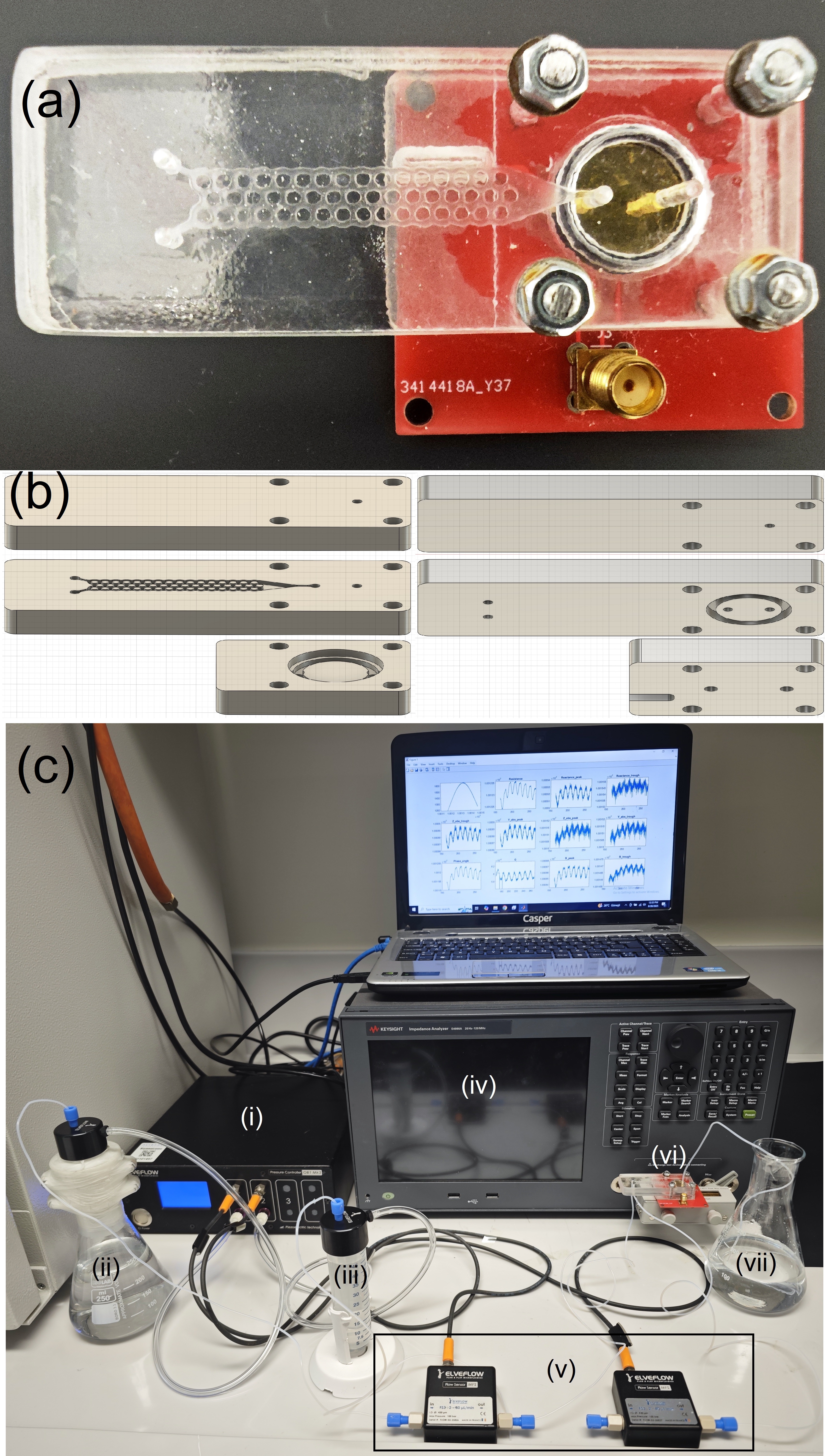}}
\caption{(a) QCM biosensor, Flowcell and Passive mixer. (b) Machined faces of passive mixer. Left: Top layers Right: Bottom layers. (c) Flow setup and experimental components.}
\label{fig1}
\end{figure}

\subsection{Impedance measurements and curve fittings}
\subsubsection{Impedance Measurements}
\begin{figure}
\centerline{\includegraphics[width=0.8\textwidth]{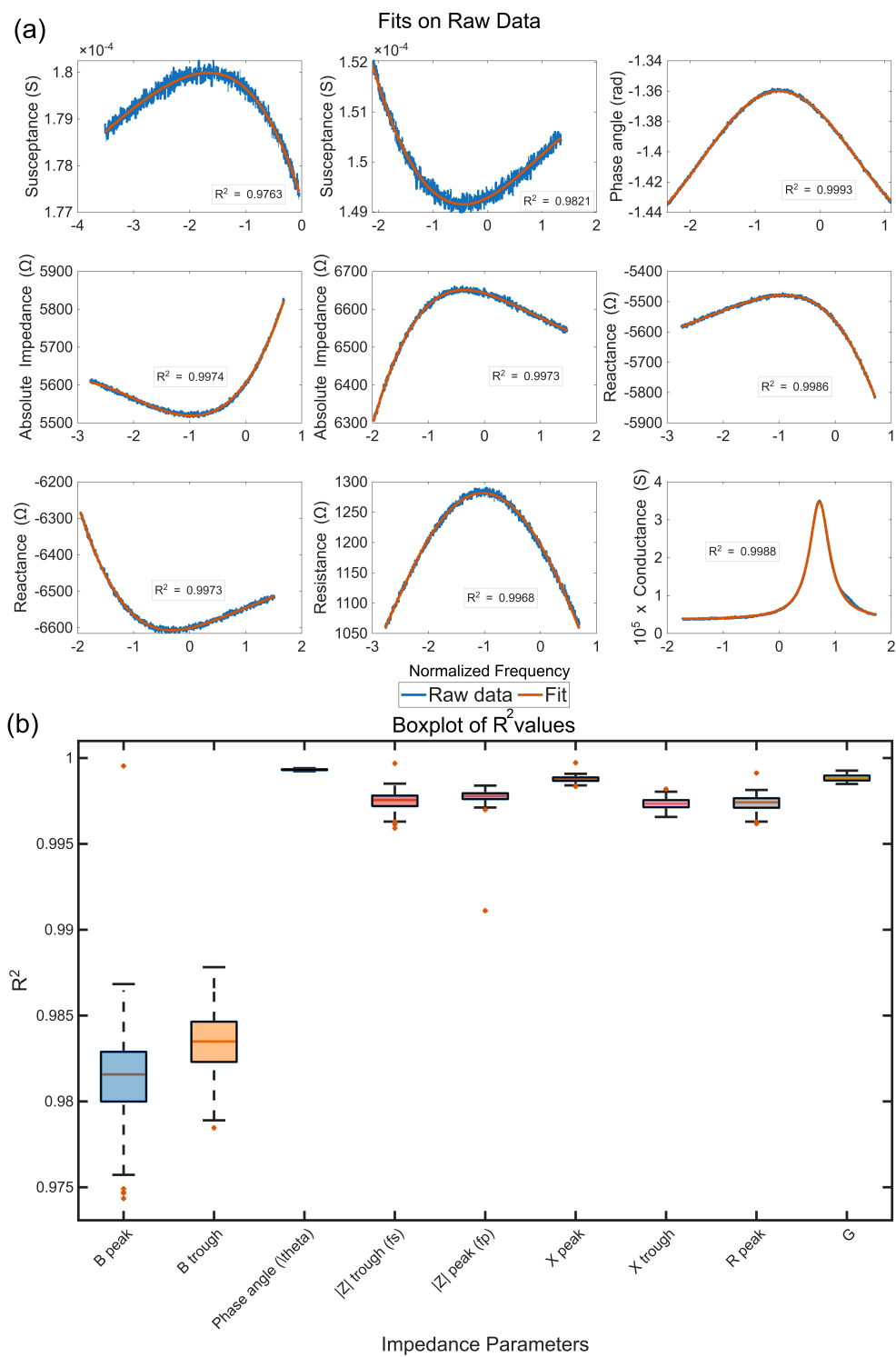}}
\caption{(a) An example of impedance spectrum around peaks and troughs. (b) Box plots of $R^2$-values for curve fittings on experimental data giving Impedance spectrum of measurements}
\label{fig2}
\end{figure}

Impedance spectra were collected around the fundamental resonance $(\sim 10 MHz)$ with 1000 points per sweep and no on-instrument averaging. Each resonance feature was acquired in its own narrow moving frequency window, centered on the feature for all measurements except conductance. The window sizes for each parameter can be find on Table 1. Real-time feature tracking and recentering were applied to all peaks/troughs except $G$ using a peak-monitoring routine based on multiple-parabola resonance peak fitting \cite{Kirimli_2014}. The conductance peak was instead measured in a fixed 50 kHz window to guarantee that both tails of the line shape remained within the sweep even under slow drift.
Fitting-window selection was predefined and feature-specific to preserve both central and wing information of each resonance observable; the complete windows are explicitly reported in Table~\ref{tab:impedance_windows} and representative wing coverage is visible in Figure~\ref{fig2}a.

\begin{table}[t]
\caption{Resonance features and sweep windows (1000 points, no averaging).}
\label{tab:impedance_windows}
\centering
\begin{tabular}{ccc}
\hline
Feature & Window Size & Center Frequency \\
\hline
$B_{\text{peak}}$  & 3 kHz & 10008449 \\
$B_{\text{trough}}$  & 3 kHz & 10014600\\
$Z_{\theta}$  & 5 kHz & 10011947\\
$|Z|_{\text{trough}}$  & 5 kHz & 10008836\\
$|Z|_{\text{peak}}$  & 5 kHz & 10015060\\
$X_{\text{peak}}$  & 5 kHz & 10009210\\
$X_{\text{trough}}$  & 5 kHz & 10015444\\
$R$  & 3 kHz & 10012285\\
$G$  & 50 kHz & 10011585\\
\hline
\end{tabular}
\end{table}
\subsubsection{Curve Fittings}
Raw impedance data for each parameter was normalized for frequency only for each parameter separately by subtracting the first sweep's mean and dividing by its standard deviation to retain shifts in frequency in time after fitting. Normalized spectral data was fitted with two term gaussian except for the conductance peak which was fitted with a triple-lorentzian with constant offset (Equation (1)) given by
\begin{equation}
L(x)=\frac{a}{(x-b)^2+c^2}+d
\end{equation}
where $a, b,c$ and $d$ fitted constants. The choice of these curves makes up a total of 52 fit parameters.
Importantly, the two-term Gaussian representation used for non-conductance features is a phenomenological decomposition of a \emph{single} asymmetric resonance envelope (left-/right-wing dominant components) and is not interpreted as two distinct physical resonances or modal coupling.
 Fitting was done using lower and upper bounds which were determined by unbounded fitting all 681 sweeps for each parameter and selecting $2^{nd}$ and $98^{th}$ percentile of for each fit parameter as lower and upper bounds to prevent overfitting (An unbounded fitting example may be found in Supplemental Information Section S.2). 9 fits in $B_{peak}$ with an $R^2$ less than 0.95 is omitted from the dataset. A sample of fitted curves and raw impedance spectra is shown in Figure 2a, and a boxplot of all the $R^2$ values is plotted in Figure 2b.   
\subsection{Exploratory data analysis and regression models}
\subsubsection{Data Set}
In constructing the dataset, each non-conductance resonance feature was fitted with a two-term Gaussian model, and a heuristic ordering was applied to the fitted components to ensure consistency across spectra. Specifically, the component with the larger fitted amplitude was designated as Gaussian 1 and its parameters were recorded with the suffix “1,” while the remaining component was designated as Gaussian 2 with the suffix “2.” The parameters from these two Gaussians, collected over all non-conductance features, yielded 48 input variables in total. The conductance peak was fitted separately with a Lorentzian function, and its four parameters were appended to the feature vector, bringing the total to 52 input parameters per one sweep of each spectrum. The supervised target (output) associated with each 52-parameter vector was the corresponding glycerol concentration at the time of acquisition.
\subsection{Outlier Removal}
Outlier detection aims to identify and eliminate anomalous data points that deviate substantially from the general structure of a dataset. In experimental measurements, such deviations may arise from random noise, instrumentation errors, or uncontrolled environmental factors. However, distinguishing between genuine noise and boundary-region behavior is nontrivial, as excessive filtering may lead to the loss of informative edge cases that represent physically meaningful system limits. Therefore, a balance must be maintained between preserving boundary behaviors and removing spurious noise that could distort downstream statistical or machine-learning analyses.

To establish this balance, the elbow method was employed to determine threshold values for each outlier detection algorithm. By plotting the number of flagged instances against a percentile or sensitivity parameter, the inflection (“elbow”) point reveals a natural transition between stable and unstable classifications. This approach provides an objective and reproducible threshold selection criterion, minimizing arbitrary parameter tuning.

Parameter bounding and omission of low-quality fits ($R^2<0.95$) effectively serve as a first-level outlier removal step, ensuring that only statistically reliable curve-fitting results are included in subsequent analyses.

In this study, outliers were handled in two stages. First, curve-fitting quality control (bounded parameterization and omission of low-quality fits, $R^2<0.95$) ensured that only statistically reliable descriptors entered downstream analysis. Second, multivariate anomaly scores were computed using three complementary detectors (local distance-based outlier factor, Isolation Forest, and Mahalanobis distance with shrinkage covariance). Scores were normalized and combined via a soft-consensus score, and samples exceeding a robust threshold were excluded. Full definitions, hyperparameters, and threshold selection procedures are provided in the Supporting Information (Section~S.5).

\subsection{Heatmap of Correlations}
A correlation heat map was constructed to evaluate pairwise linear relationships among all fitted parameters. Correlation analysis quantifies the degree to which two variables vary together, with Pearson’s correlation coefficient ranging between –1 and +1, indicating the strength and direction of linear association. The heat map provides a compact, visual representation of these interdependencies, facilitating the identification of clusters of parameters that vary in a similar manner or exhibit inverse trends. This step serves both as an exploratory diagnostic to assess data coherence and as a guide for subsequent feature selection and dimensionality-reduction analyses.

\subsection{Feature Importance Analysis}
To determine the most informative parameters for predicting glycerol concentration, feature-ranking was performed using mutual information (MI)–based minimum-redundancy maximum-relevance (mRMR) analysis. Feature importance quantifies the explanatory contribution of each variable to the target response, helping to identify those parameters that convey the most useful information while minimizing redundancy among correlated features.

The mutual information (MI) between a feature $X_i$ and the target variable $Y$ measures the reduction in uncertainty about $Y$ when $X_i$ is known. Unlike linear correlation coefficients, MI can capture both linear and nonlinear dependencies. It is formally defined as $$I(X_i,Y)=\int\int p(x_i,y) log\frac{p(x_i,y)}{p(x_i)p(y)}dx_i\,dy$$ where $p(x_i,y)$ is the joint probability density function and $p(x_i)$ and $p(y)$ are the corresponding marginals. A higher MI value indicates a stronger statistical dependence between the feature and the target variable. 

To ensure that the selected features are not only relevant but also non-redundant, the mRMR algorithm was applied. The mRMR criterion seeks to maximise the relevance of the selected feature subset 
S to the target variable while minimising mutual information among features within the subset: $$\max\limits_{S}\left[\frac1{|S|}\sum\limits_{x_i\in S}I(x_i,Y)-\frac{1}{|S|^2}\sum\limits_{x_i,x_j\in S} I(x_i,x_j)\right].$$ This formulation ensures that the resulting subset of parameters carries the maximum possible information about the response variable without duplication of information between features.

In this study, MI was employed as the primary dependency measure within the mRMR framework, as it provides a deeper and more comprehensive assessment of feature relevance than linear correlation metrics. The resulting ranked list of features was subsequently used to evaluate regression model performance as a function of the number of included parameters.

\subsection{Regression Modeling}
Regression models were employed to quantitatively assess the effectiveness of the extracted and ranked features in predicting glycerol concentration. The objective was to evaluate the predictive utility of impedance line-shape descriptors and the effect of redundancy-aware feature ranking under a consistent cross-validation protocol, rather than to claim a universally optimal model. To provide a balanced view of model interpretability and nonlinear learning capacity, three distinct classes of regression algorithms were considered.
\subsubsection{Linear Models}
The Lasso and Elastic Net regressions were implemented to evaluate the predictive power of sparse linear representations. Lasso regression imposes an $L_1$-norm penalty on the regression coefficients, forcing some weights to zero and thereby performing both regularisation and feature selection. Elastic Net combines the $L_1$ and $L_2$-norm penalties, providing a compromise between the sparsity of Lasso and the stability of Ridge regression. These models serve as interpretable baselines that reveal whether a linear combination of selected Gaussian- and Lorentzian-fit parameters can adequately describe concentration variations.
\subsubsection{Operator-Theoretic Model}
The Support Vector Regressor (SVR) was used as a kernel-based approach that projects the input space into a higher-dimensional feature space where linear regression is performed within a tolerance margin $\varepsilon$. SVR is derived from operator theory and optimization principles, where the regression function is constructed by minimizing structural risk rather than empirical error. This formulation allows the model to capture nonlinear trends while maintaining good generalization performance and resistance to overfitting.
\subsubsection{Algorithmic Ensemble Models}
To further explore nonlinear relationships and variable interactions, several ensemble-based tree models were trained: Random Forest (RF), Gradient Boosting Regressor (GBR), Extreme Gradient Boosting (XGBoost), and CatBoost.
Random Forest aggregates predictions of multiple decorrelated decision trees to reduce variance and enhance robustness. Gradient Boosting and its optimized variants (XGBoost, CatBoost) sequentially build trees where each tree corrects the errors of the previous ensemble. CatBoost, in particular, improves handling of feature interactions and prevents overfitting through ordered boosting and internal regularization.
These algorithmic models provide complementary insights into complex, nonlinear dependencies that may not be captured by purely linear approaches.

All models were trained under a unified five-fold cross-validation framework to ensure consistency and comparability of results. Performance metrics including the coefficient of determination ($R^2$), were computed for both training and validation sets. The resulting performance trends as a function of feature subset size were used to validate the reliability and interpretability of the feature-importance analysis.

All analyses were performed on the same post-QC dataset (N = 648 sweeps), using a fixed random seed for fold assignment. Feature ranking (mRMR) was computed within each training fold and applied to the held-out fold to avoid information leakage. Models were evaluated using $R^2$ and RMSE on validation folds; RMSE is additionally reported for comparison against the Kanazawa baseline.

\subsection{Simulated Resonance Behavior Analysis Using the Butterworth--Van Dyke Equivalent Circuit}

To support interpretation of the impedance line-shape descriptors extracted from experimental spectra, we performed an illustrative simulation using the Butterworth--Van Dyke (BvD) equivalent circuit. The BvD model provides a physically grounded representation of a quartz resonator through a motional branch \((R_m, L_m, C_m)\) in parallel with the static capacitance \((C_0)\). By imposing controlled perturbations in \(R_m\) and \(L_m\), the simulation generates corresponding, physically consistent changes in the resonance response (e.g., resonance frequency, conductance, and phase), enabling a direct check of how the fitted Gaussian/Lorentzian descriptor set responds to known circuit-level variations. Importantly, this simulation is used to enhance interpretability of the extracted features and to provide a controlled reference for trend analysis; it is not intended as a comprehensive surrogate for the full experimental loading physics.

Specifically, we simulated the experiment using the crystal parameters \(R_m = 5~\Omega\), \(L_m = 9 \times 10^{-3}~\mathrm{H}\), \(C_m = 28 \times 10^{-15}~\mathrm{F}\), and \(C_0 = 5 \times 10^{-12}~\mathrm{F}\). A sinusoidal loading pattern was emulated by generating a \(50~\mathrm{Hz}\) peak-to-peak shift in the phase-angle peak position over one period comprising 100 sweeps. This target shift was achieved by determining sweep-wise \(L_m\) and \(R_m\) updates through minimization of the peak-position error, using the same real-time peak-tracking routine applied to the experimental data. The same window sizes and data resolution were maintained as in the experimental dataset. \(R_m\) and \(L_m\) were re-estimated after each round of nine sweeps, covering all troughs and peaks of the studied impedance parameters. The simulated spectra were then processed with the same downstream pipeline (feature extraction and filtering) up to the bivariate analysis stage, enabling like-for-like comparison of descriptor behavior between simulated perturbations and experimental observations.
\subsection{Classical Kanazawa Model based on Impedance}
Kanazawa based model predictions are calculated using half-bandwidth, $\Gamma$, which for Newtonian fluids shifts equally with $f_{r}$, fundamental resonance frequency\cite{Johannsmann2021_Sensors_Review}. The calculations involves peak position and baseline determination of the conductance, and subsequent $\Gamma$ detection as explained in detail before \cite{Kirimli_2025}. The viscosity values calculated from the impedance data was then converted into data oriented concentration as explained before and compared with the actual glycerol concentrations. 

\section{Results and Discussion}
\subsection{Curve fitting and parametric representation}
The impedance spectrum around peak and troughs are not symmetrical as can be seen figure 2a. In order not to loose this vital information functions retaining important parameters such as peak height, width and position were chosen to fit with high $R^2$ with least amount of parameters. Asymmetry in gaussians require at least 2 terms. Fitting 2 term gaussians caused overfitting as shown in supplemental information, Section S.2. Centers of individial gaussians converge to outlier frequencies for a minimal gain in $R^2$ increase making a pattern with changing with concentration difficult to emerge. To prevent this an unbounded fitting was used to determine boundaries for fit values as described which lowered the $R^2$ values, as can be seen in Figure 2b the lowest $R^2$ value was 0.9743. 

Each impedance spectrum was reduced to a compact, physically meaningful set of line-shape descriptors. For all parameters except conductance G, the spectral neighbourhood is modeled with a two-term Gaussian mixture, capturing peak intensity, location, and width while remaining robust to minor asymmetries and baseline drift. The conductance peak, which is theoretically Lorentzian from the BvD model \cite{Arnau2008_Sensors}, was fitted with a triple-Lorentzian plus constant offset to accommodate shoulders and heavy tails observed experimentally. Bounds on fit coefficients were derived from percentile statistics over all sweeps to prevent overfitting and ensure reproducibility. This procedure yields a 52-feature representation per one sweep of each spectrum that preserves the core resonance information in a form suitable for multivariate analysis. We selected Gaussian and Lorentzian parameterizations, although third-order polynomial models also achieved high-$R^2$ fits to the line shape. This choice preserves interpretability by yielding direct descriptors of resonance morphology—peak position, peak amplitude, and width (FWHM-related quantities)—whereas polynomial coefficients are less physically transparent.

Figure 2b summarizes goodness-of-fit across parameters: with the exception of Susceptance, all $R^2$ values exceed 0.995, indicating that the chosen parametric forms closely track the measured spectra. The slightly lower $R^2 (\geq 0.97)$ for Susceptance likely reflects its higher noise sensitivity and broader low-curvature region, but residuals remain small and structureless relative to signal magnitude. Combined with the prior exclusion of low-fidelity fits and percentile-based coefficient bounds, these results confirm that the fitted features provide a faithful and stable surrogate of the raw spectra, suitable for downstream correlation analysis, feature ranking, and regression.

\subsection{Outlier detection}
\subsubsection{Ordering in the Venn diagram}
Figure 3a shows that Isolation Forest (IF) removed the most points (62), Local Distance–Based Outlier Factor (LDOF) fewer (17), and Mahalanobis distance (MD) the fewest (2), with MD a subset of the other two. This ordering is consistent with each method’s sensitivity: IF isolates observations by random partitions and therefore readily flags sparse tails and small peripheral clusters (higher recall); LDOF emphasizes local density contrast and is stricter when a point lies near a cohesive neighborhood (moderate recall); MD assumes an approximately elliptical global structure with shrinkage covariance and typically marks only the most extreme, directionally deviant points (highest precision, lowest count). Hence the empirical pattern $IF > LDOF > MD$, and MD points tend to fall within the sets detected by IF/LDOF.
\begin{figure}
\centerline{\includegraphics[width=0.9\textwidth]{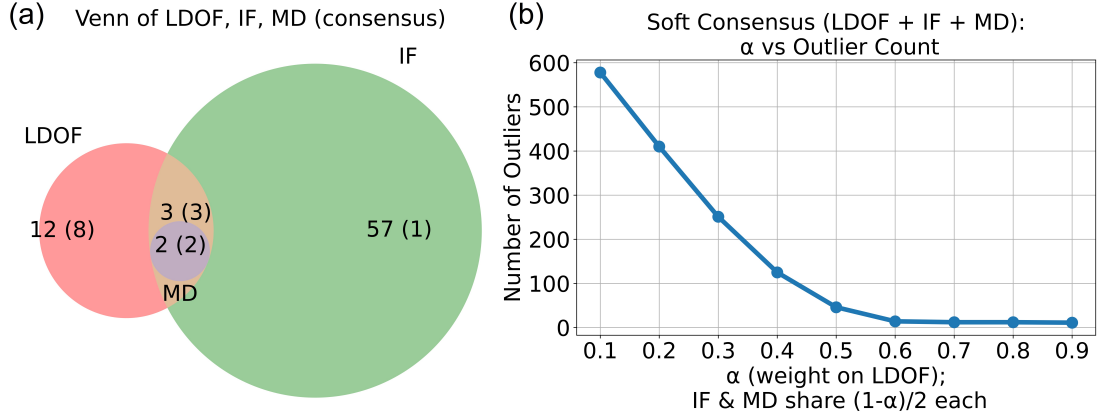}}
\caption{(a) Venn diagram of outlier points labelled independently by different (LDOF: local distance-based outlier factor, IF: Isolation forest, MD: Mahalanobis distance) algorithms. First number is the number of data points labelled by each method. The number in paranthesis denotes the number of those still labelled as outlier by consensus method (b) Graph of number of outliers consensus method sketched with respect to hyperparameter $\alpha$}
\label{fig3}
\end{figure}
\subsubsection{Soft consensus method}
None of the above three methods lead to an elbow graph which is an indication that individual outlier detection methods do not catch the boundary behavior (Details may be found in Supplemental Information Section S.5). To obtain a unified decision, a soft consensus ensemble method is employed that fuses continuous anomaly scores rather than binary votes. After robust normalization to $[0,1]$ a weighted score is computed and an outlier is flagged when $S\geq \tau$ (here $\tau=0.5$ is the threshold). Because soft consensus aggregates sub-threshold but consistent evidence across detectors, it can identify points that do not appear in any single method’s binary set—hence additional outliers “outside” the Venn diagram. In contrast, hard consensus operates on binary masks only: a union rule (flag if any method flags) maximises recall, while an intersection rule (flag only if all methods agree) maximises precision but may miss borderline cases; neither leverages graded evidence.

\subsubsection{Elbow Selection of the Weight Alpha}
The weights are parametrized as $w_{LDOF}=\alpha$ and $w_{IF}=W_{MD}=(1-\alpha)/2$. Plotting the number of flagged outliers versus $\alpha$ produces a characteristic elbow curve, seen in Figure 3b: large changes in count for small $\alpha$ (dominated by IF/MD), followed by a region of diminishing returns as $\alpha$ increases (greater emphasis on LDOF’s local-density view). $\alpha$ is selected at the elbow—where the slope drops sharply—because it balances noise removal against boundary preservation and yields stable downstream behavior. In our study, this criterion led to $\alpha=0.6$.
\begin{figure}
\centering
\includegraphics[width=\textwidth]{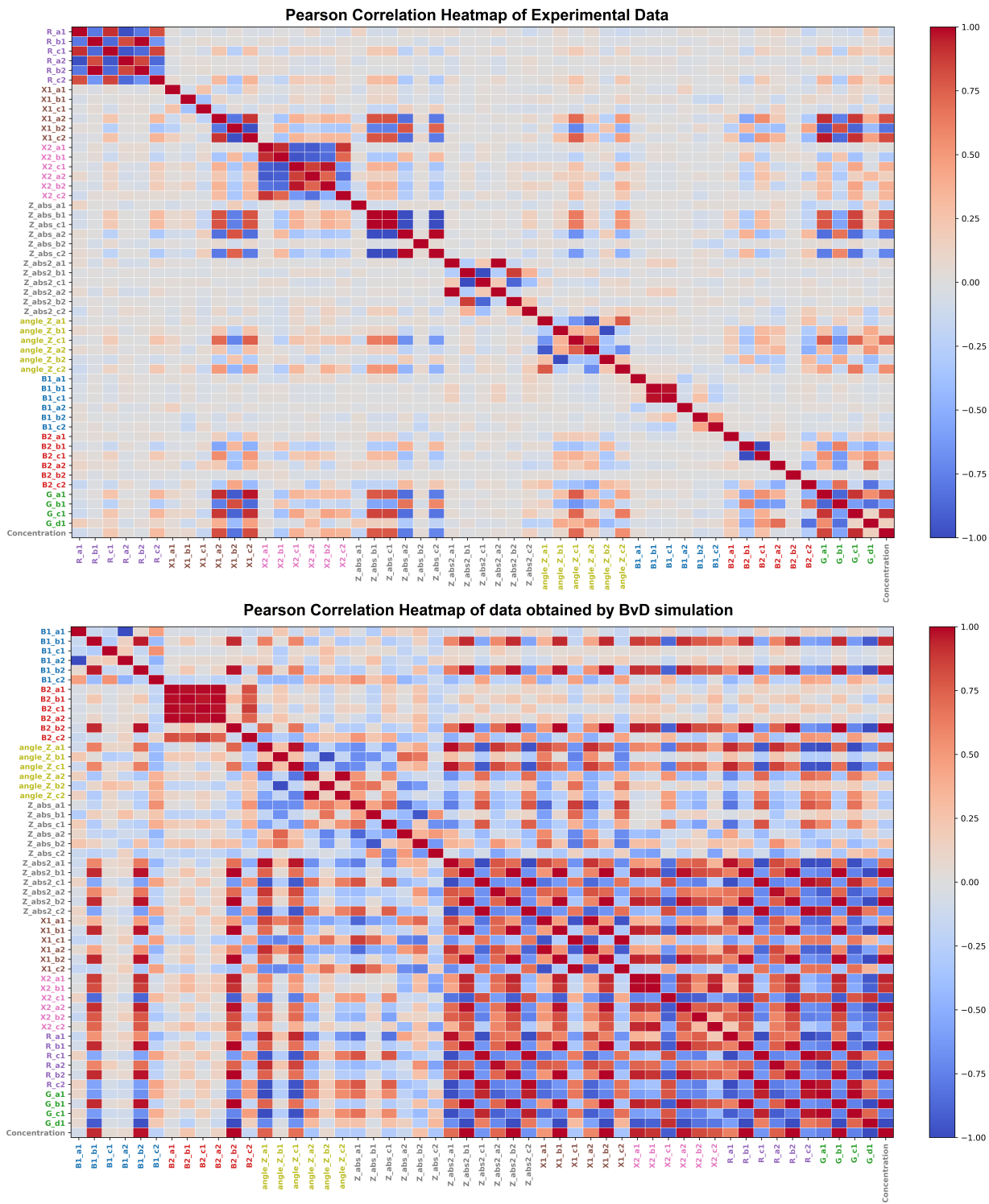}
\caption{(a) Heatmap of Pearson correlation values computed for experimental data (b) Heatmap of Pearson correlation computed for BvD simulation results}
\label{fig4}
\end{figure}

\begin{figure}
\centerline{\includegraphics[width=0.6\textwidth]{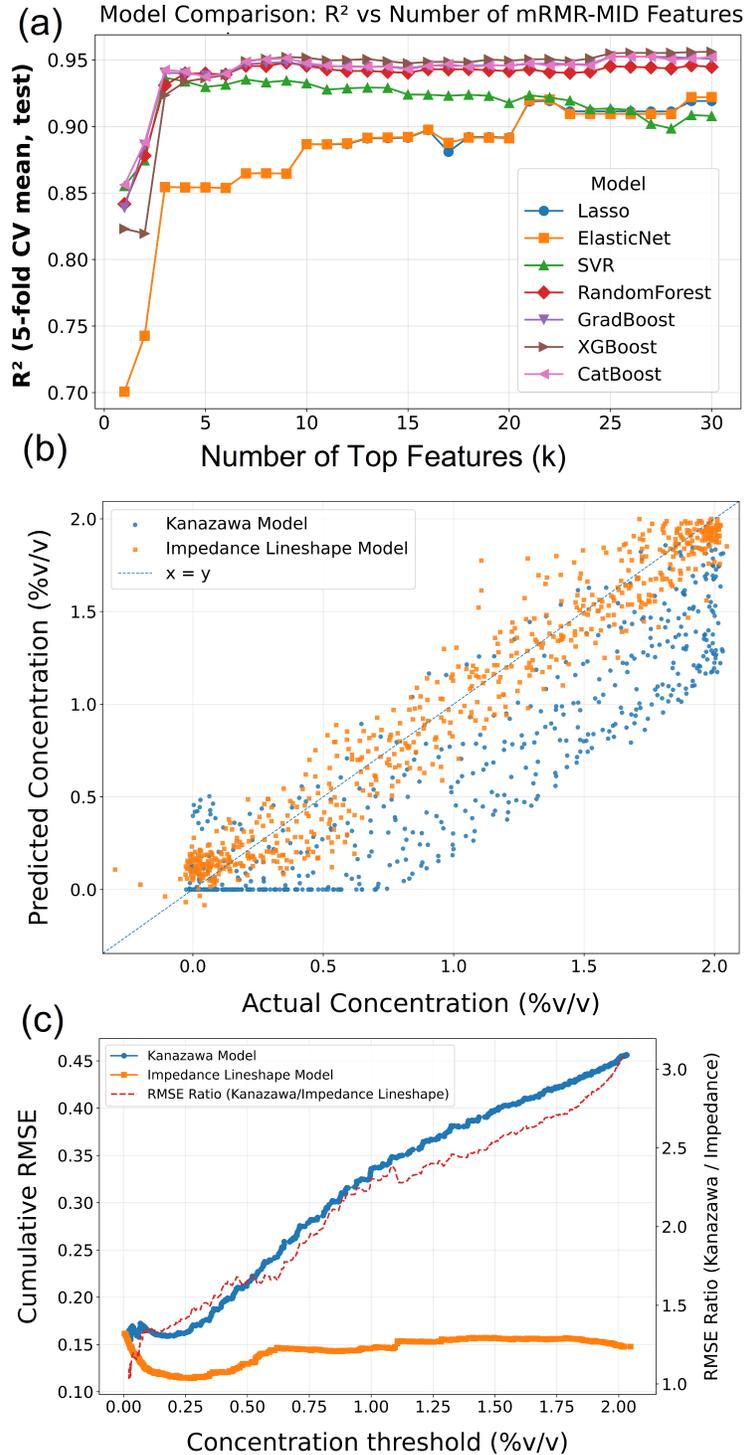}}
\caption{(a) Graph of $R^2$ values with respect to number of mRMR-MID sorted top features employed in training for each regression model. (b)Graph of actual concentration vs. predicted concentrations for Kanazawa model and Impedance based XGBoost regression model. (c)Cumulative RMSE values for the Kanazawa and impedance line-shape models are plotted against increasing concentration on the left y-axis, while their RMSE ratio (Kanazawa/impedance line-shape) is shown on the right y-axis.}
\label{fig5}
\end{figure}

\subsection{Correlations}
Figure 4 shows a heat map of the correlation matrix where the matrix has unity diagonal due to perfect self-correlation. When focused around the diagonal, for $R, X_{trough}$ both the Gaussian terms' fit parameters and $G$, all fit parameters are highly correlated. $\Theta_{Z}, |Z|_{peak}$ and $|Z|_{trough}$ shows moderate correlation and $X_{peak}$ only the second Gaussian term shows correlation. $B_{peak}$ shows correlation only for the parameters b(peak position) and c(FWHM=$4.c^2.ln2$)  for both gaussian terms and $B_{trough}$ does the same for only the first Gaussian term. Peaks and troughs correspond to the series and parallel resonance frequencies of the Butterworth -Van Dyke equivalent circuit model (BvD). \[f_s=\frac1{2\pi\sqrt{L_mC_m}}\qquad \text{and }\qquad f_p=f_s^{lossless}\sqrt{\frac1{\left(1+\frac{C_m}{C_0}\right)}}\] where $C_m\ll C_0$. Although, BvD is a simplified model and variations of the BvD model and computational studies on the BvD itself is studied extensively in the literature to better represent the behavior of QCM under load from Newtonian fluids. \cite{Martin1991,Muramatsu1988,Huang2017,Arnau2008_Sensors}. As both fs and fp shifts to lower frequencies according to BvD model upon changes in motional inductance ($L_m$) and motinal resistance ($R_m$), due to bulk fluid immersion, more of the parameters were expected to be correlated.  Although the asymmetrical impedance peaks were captured fully in simulation as shown in Supplemental Information Section S.3, bivariate analysis of the simulation results of the BvD model shows more of the parameters to be correlated and the pattern of those correlations from stronger to weaker is also different. These results suggest there is a more complicated underlying mechanism in QCM's response to changes in viscosity of the Newtonian fluid that is not fully captured by the simplified BvD formulation used here, suggesting additional contributions beyond the modeled $R_m/L_m$ perturbations. Moreover, although in reality all the peak positions showed high correlation with the concentration with  sinusoidal patterns(Figures may be found in Supplementl Information Section S.9), fit parameters in this study captured more nonlinear behavior when combined with regression models predicted the changes in concentration with significantly higher accuracies.

\subsection{Parameter importance}
A linear dimensionality analysis of the data set is applied using principal component analysis (PCA). Results may be found in Supplemental Information Section S.6. Table 2 shows the importance of parameters in explaining the variations in the concentration. Other statistical methods to determine parameter information may be found in Supplemental Information Section S.4. The most important parameters is the height of the dissipation peak, followed by the second Gaussian term of the phase angle. Phase angle peak is situated in between the $f_s$ and $f_p$ peaks and as both frequencies shift phase angle peak, shifts. The results indicate the tail of the peak closer to $f_s$ is more responsive to concentration changes. The peak position of the conductance is the third parameter, followed by reactance trough's second Gaussian term's width. In the top 10 most important parameters 3 out of 4 terms of the conductance peak is followed by 3 terms of the phase angle covering both of its tails, which is followed by 2 terms of $|Z|$'s $f_s$ trough both tails' amplitudes. It is important to note that, the sorting algorithm employed here increases the convergence of the regression algorithm with as little parameters as possible by penalizing unsorted parameters with how redundant they are with respect to already selected parameters in explaining concenumberntration. This means if one has to pick a parameter out of this ranking, a parameter closer to the end may perform quite well alone. (The bar graphs of mRMR-MID scores may be found in Supplemantal Information Section S.7)

\begin{table}[!t]
\centering
\small
\setlength{\tabcolsep}{4pt}
\renewcommand{\arraystretch}{1.8}

\begin{threeparttable}
\caption{\parbox{0.6\linewidth}{Rank-sorted fit parameters. Columns indicate contiguous rank ranges; rank increases top-to-bottom within each column and then continues in the next column.}}
\label{tab:ranked_params_with_legend_overline}

\begin{tabular}{l l l l l}
\textbf{Ranks 1--11} & \textbf{Ranks 12--22} & \textbf{Ranks 23--33} & \textbf{Ranks 34--44} & \textbf{Ranks 45--52} \\

$G^{a}$                            & $\xb{Z_{\Theta}^{b1}}$                 & $\xbu{|Z|_{peak}^{a2}}$              & \underline{$X_{trough}^{a1}$}        & \underline{$X_{trough}^{c1}$} \\
$\xb{Z_{\Theta}^{c2}}$             & $\xbiu{B_{trough}^{c1}}$               & $\xbi{|Z|_{trough}^{b2}}$            & \emph{$X_{peak}^{a1}$}               & \emph{\underline{$B_{peak}^{b2}$}} \\
$G^{b}$                            & \emph{$X_{peak}^{b2}$}                 & $\xbi{|Z|_{trough}^{c1}}$            & $\xbiu{B_{trough}^{a1}}$             & \emph{\underline{$B_{peak}^{a1}$}} \\
\underline{$X_{trough}^{c2}$}      & $\xbu{|Z|_{peak}^{b2}}$                & $\xbu{|Z|_{peak}^{b1}}$              & \emph{\underline{$B_{peak}^{c2}$}}   & $\xb{Z_{\Theta}^{a2}}$ \\
$\xb{Z_{\Theta}^{b2}}$             & $\xbi{|Z|_{trough}^{c2}}$              & \emph{$X_{peak}^{c1}$}               & $\xbi{|Z|_{trough}^{b1}}$            & $\overline{\mathrm{R^{a1}}}$ \\
\emph{$X_{peak}^{a2}$}             & $\xbiu{B_{trough}^{b2}}$               & $\overline{\mathrm{R^{c2}}}$         & $\overline{\mathrm{R^{c1}}}$         & \emph{\underline{$B_{peak}^{b1}$}} \\
$G^{c}$                            & \emph{$X_{peak}^{c2}$}                 & $\xbiu{B_{trough}^{a2}}$             & $\xbu{|Z|_{peak}^{c1}}$              & \underline{$X_{trough}^{a2}$} \\
$\xbi{|Z|_{trough}^{a1}}$          & \emph{\underline{$B_{peak}^{a2}$}}     & \underline{$X_{trough}^{b2}$}        & $\xb{Z_{\Theta}^{a1}}$               & $\overline{\mathrm{R^{b1}}}$ \\
$\xbi{|Z|_{trough}^{a2}}$          & \emph{$X_{peak}^{b1}$}                 & $\xbiu{B_{trough}^{b1}}$             & \emph{\underline{$B_{peak}^{c1}$}}   &  \\
$\xb{Z_{\Theta}^{c1}}$             & $\xbiu{B_{trough}^{c2}}$               & $\xbu{|Z|_{peak}^{a1}}$              & $\xbu{|Z|_{peak}^{c2}}$              &  \\
$\overline{\mathrm{R^{b2}}}$       & \underline{$X_{trough}^{b1}$}          & $G^{d}$                               & $\overline{\mathrm{R^{a2}}}$         &  \\

\end{tabular}

\begin{tablenotes}[flushleft]
\footnotesize
\item \textbf{Legend (font style $\rightarrow$ parameter family):}
Normal $\rightarrow$ $G$;\;
Bold $\rightarrow$ $Z_{\Theta}$;\;
Italic $\rightarrow$ $X_{peak}$;\;
Underline $\rightarrow$ $X_{trough}$;\;
Bold+Italic $\rightarrow$ $|Z|_{trough}$;\;
Bold+Underline $\rightarrow$ $|Z|_{peak}$;\;
Italic+Underline $\rightarrow$ $B_{peak}$;\;
Bold+Italic+Underline $\rightarrow$ $B_{trough}$;\;
$\overline{\mathrm{Sample}}$ $\rightarrow$ $R$.
Exponents represent fit terms; Gaussian fits are two-termed \((a1,b1,c1,a2,b2,c2)\); a triple-Lorentzian with constant offset was fit on $G$ \((a,b,c,d)\).
\end{tablenotes}

\end{threeparttable}
\end{table}

\subsection{Regression Models}
To evaluate the predictive value of the fitted line-shape parameters and to verify the benefit of feature selection, a set of regression models (Lasso, Elastic Net, SVR, Random Forest, Gradient Boosting, XGBoost, CatBoost) are fitted under a unified 5-fold cross-validation protocol and assessed performance with $R^2$. For each model, performance of the model as a function of the top-
k features ranked by MI-based mRMR. The resulting curves show that accuracy rises steeply with a small number of ranked features and then plateaus, indicating that a compact, low-redundancy subset captures most of the predictive signal. Compared with classical impedance descriptors (e.g., single peak position/height or a small hand-crafted set), the mRMR-guided subsets systematically achieve higher $R^2$, demonstrating that the parametric line-shape features (centers, widths, amplitudes across families, including the multi-peak conductance fit) provide richer and more stable information about concentration. Taken together, these results confirm that the proposed features are genuinely informative for prediction, and redundancy-aware selection improves both parsimony and generalisation without sacrificing interpretability. 

In Figure 5a, the SVR’s test performance declines as more features are added because the additional variables beyond the top mRMR set contribute little signal but increase dimensional noise and redundancy. With an RBF kernel, Euclidean distances become less discriminative in higher dimensions (“distance dilution”), so fixed hyperparameters no longer match the effective smoothness and margin needed as k grows. Consequently, the model over-smooths or overfits along irrelevant directions, reducing generalization. $R^2$, Mean absolute error (MAE) and Root mean square error (RMSE) graphs for each model may be found in Supplemental Information Section S.8. Best performing regression model was XGBoost with 30 top features. Figure 5b and 5c compares the concentration prediction of the Kanazawa's models with the best performing impedance based model. Predicted concentrations were plotted against actual concentration values as calculated from flow sensors in Figure 5b. Other regression models' prediction were plotted in Supplemental Information Section S.10. All the models at their best performing top features performed with $RMSE=0.456$, significantly  outperforming Kanazawa based model predictions $RMSE=0.148$ value, calculated using half-bandwidth, $\Gamma$, which for Newtonian fluids shifts equally with $f_{r}$, fundamental resonance frequency. Figure 5b also shows the predictions at low concentrations were significantly better than Kanazawa model. To better visualize this, cumulative $RMSE$ values were plotted  against threshold concentrations, where $RMSE$ was calculated for concentrations lower than this threshold concentration as shown in Figure 5c (left y-axis). The ratio of cumulative RMSE's were also plotted to compare the impedance lineshape model with Kanazawa model in figure 5c (right y-axis). Best performing regression model significantly outperformed Kanazawa model, showing same level of prediction even at low concentrations between 0.25-1.00\%.

\section{Conclusion}
Impedance-resolved QCM, integrated with a passive microfluidic mixer and a redundancy-aware machine learning pipeline, was shown to enable accurate, real-time prediction of small composition changes in the $0$--$2\%$ (v/v) glycerol range. By converting each sweep of nine spectra into 52 physically interpretable line-shape features and applying consensus outlier handling followed by mRMR (mutual information difference) ranking, compact subsets dominated by conductance, phase, and reactance descriptors drove regressors to $R^{2}>0.95$. Across all models, the impedance line-shape approach consistently exceeded baselines that rely only on $\Delta f$ or $\Delta D$, with the most pronounced gains at low \% (v/v). These results indicate that preserving the full resonance morphology—beyond two-metric readouts—unlocks robust, information-rich signals for concentration prediction. More generally, the study provides an AI-ready pattern for converting high-resolution spectral sensor readouts into compact, interpretable features coupled with redundancy control and cross-validated regression.

From an instrumentation perspective, these findings suggest that QCM workflows constrained to $\Delta f/\Delta D$-only endpoints can become an information bottleneck for inference tasks. When prediction accuracy is the primary objective, restricting acquisition and analysis to two scalar readouts may be suboptimal compared with multi-observable, impedance-resolved line-shape processing.

The workflow is model-agnostic and hardware-light, making it readily extensible to additional analytes, viscoelastic films, higher overtones, and other piezoelectric resonators. While this study focuses on Newtonian glycerol–water mixtures under controlled flow, the same impedance-resolved feature extraction and learning framework is directly applicable to broader analytes and more complex fluids, provided that appropriate calibration data are available.

%%%%%%%%%%%%%%%%%%%%%%%%%%%%%%%%%%%%%%%%%%%%%%%%%%%%%%%%%%%%%%%%%%%%%

%%%%%%%%%%%%%%%%%%%%%%%%%%%%%%%%%%%%%%%%%%%%%%%%%%%%%%%%%%%%%%%%%%%%%
% Supporting Information statement (required if SI is provided)
%%%%%%%%%%%%%%%%%%%%%%%%%%%%%%%%%%%%%%%%%%%%%%%%%%%%%%%%%%%%%%%%%%%%%
\section*{Supporting Information}
Additional experimental details, supplementary figures, and supplementary tables are provided in the supplementary PDF distributed as an ancillary file with the arXiv submission.

%%%%%%%%%%%%%%%%%%%%%%%%%%%%%%%%%%%%%%%%%%%%%%%%%%%%%%%%%%%%%%%%%%%%%
% Author information (optional but commonly included in ACS manuscripts)
%%%%%%%%%%%%%%%%%%%%%%%%%%%%%%%%%%%%%%%%%%%%%%%%%%%%%%%%%%%%%%%%%%%%%
\section{CRediT authorship contribution statement}
C.K.:conceptualization, resources, writing-original draft, supervision, project administration, funding acquisition. E.E.:software, data curation, writing-review and editing,supervision. S.Y.T.:investigation. 
\section{Declaration of Competing Interest}
The authors declare that they have no known competing financial interests or personal relationships that could have appeared to influence the work reported in this paper.
\section{Acknowledgements}
This work is supported partially by the Scientific and Technological Research Council of Türkiye, grant number:122E166. We would also like to thank Boğaziçi University Life Sciences and Technologies Application and Research Center for lending their microfluidic pump system. 
\section{Data Availability}
Data will be made available on request.


\bibliographystyle{plainnat}
\FloatBarrier
\clearpage
\bibliography{references}

\end{document}